\documentclass[twocolumn,aps]{revtex4}

\begin{document}
\draft
\title{Comment on  ``Decoy State Quantum Key Distribution''}
\author{Xiang-Bin Wang\thanks{Email address: wang@qci.jst.go.jp}\\
IMAI Quantum Computation and Information Project,
ERATO, JST,
Daini Hongo White Bldg. 201, \\5-28-3, Hongo, Bunkyo,
Tokyo 133-0033, Japan}
\begin{abstract} 
The main claim by H.K. Lo et al that they have for the first time
made the decoy-state method efficiently work in practice is inappropriate.
We show that, prior to our work, actually (and obviously) none of proposals raised by H.K. Lo et al
can really work in practice. Their main protocol requires infinite number
of different coherent states which is in principle impossible for any set-up.
Their idea of using very weak coherent state as decoy state doesn't work 
either by our detailed analysis. The idea implicitly requires an 
unreasonablly large number of pulses which needs at least 14 days to produce, if they want to do QKD
over a distance of 120-130km. 
\end{abstract}
\maketitle
The recent paper\cite{lolo}, quant-ph/0411004 by H. K. Lo et al claims that
they have {\em for the first time} made the decoy-state method efficiently useful in practice.
The paper\cite{lolo} is an extended version of their earlier results announced in a number of conferences\cite{tot,lo4}.
 We shall show that, actually, none of their proposal really works.

Their main protocol requires infinite pulses. This fact has been emphasized by H.K. Lo in a number of conferences\cite{tot}.
Here are some statements quoted from Ref.\cite{tot}(page 27 or page 18):
\begin{center} ``Idea''\end{center}
``Try  every Poisson distribution $\mu$ !''. 
``We propose that Alice switches power of her laser up and down, thus producing as
decoy states Poisson photon number distributions, $\mu'$s for {\bf all} possible values of $\mu'$s.''.
In Lo's transparency, the words ``every'' and ``all'' are highlighted.
The main protocol in their recent presentation\cite{lolo} is 
obviously the same with that in ref\cite{tot}.

Actually, the main protocol given by Lo et al\cite{tot,lolo} is even worse than the Trivial idea of using single photon source.
Trivial idea  is more feasible than Lo's main protocol: Although single photon source is difficult, it is at least
in principle possible. However, producing infinite number of coherent states is {\em in principle impossible}. 
  
In Ref\cite{lo4}, another idea by H. K. Lo is shortly stated:``On one hand, by using a vacuum as decoy state, 
Alice and Bob can verify the so called dark count rates of their detectors. On the other
hand, by using a very weak coherent pulse as decoy state, Alice and Bob can easily lower bound
the yield (channel transmittance) of single-photon pulses.''
We now show that this idea doesn't work either. 
By the idea, they need two sets of
decoy pulses: Set $Y_0$ contains $M$ vacuum pulses $|0\rangle\langle 0|$ and
set $Y_v$ contains $N$ pulses of very weak coherent state $|\mu_v\rangle\langle \mu_v|$. 
They can only observe the total counts of set $Y_0$ and the total counts of set $Y_v$.
 By that idea\cite{lo4}, to verify a meaningful lower bound of
single photon yield, $s_1$, the value $\mu_v$  must be less than channel transmittance $\eta$.
For clarity, we assume zero dark count first. 
In the normal case when there is no Eve,  $N$ decoy pulses in class $Y_v$ will cause $N(1-e^{-\eta\mu_v})$ counts.
For the security, one has no other choice but to assume the
worst case that all multi-photon pulses have caused a count. Therefore the lower bound
of single-photon counts is $N[1-e^{-\eta \mu_v}-(1-e^{-\mu_v}-\mu_v e^{-\mu_v})]=N(\eta\mu_v-\mu_v^2/2)$.
The lower bound value for $s_1$ is verified by  $s_1\ge \frac{N(\eta\mu_v-\mu_v^2/2)}{N\mu_ve^{-\mu_v}}\approx \eta -\mu_v/2$.
Therefore one has to request $\mu_v\le\eta$ here if one wants to verify 
$s_1\ge \eta/2$.
Now we consider the effect caused by dark counts.
Suppose, after observed the counts of pulses in set $Y_0$, they find that the dark count rate, $s_0=10^{-6}$
for set $Y_0$. Note that the dark count rate for set $Y_0$ and the dark count rate for set $Y_v$ can be a little bit different due to the
statistical fluctuation.    
Given $N$ pulses of state $|\mu_v\rangle$, there are $Ne^{-\mu_v}$ vacuum pulses
and $N(1-e^{-\mu_v})$ non-vacuum pulses. Alice does not know which
pulse is vacuum which pulse is non-vacuum. They can $only$ observe
the number of total counts ($n_t$) caused by $N$ decoy pulses in set $Y_v$, which is
the summation of dark counts, $n_0$, the number of single-photon counts $n_1$ and the number of
multi-photon counts, $n_m$, of those $N$ decoy pulses in set $Y_v$.
 After observed the number of total counts $n_t$, they try to estimates
$n_1$ by the formula $n_t=n_0+n_1+n_m$, with $n_0=Ns_0' e^{-\mu_v}$ and 
the worst-case assumption of $n_m= N(1-e^{-\mu_v}-\mu_v e^{-\mu_v})$.
The value $s_0'$ is the dark count rate for set $Y_v$ and the value $s_0'$
 is never known $exactly$. They only know the approximate value, $s_0'\approx s_0 =10^{-6}$. 
Consider the case $\eta=10^{-4}$. (Remark: Here the device loss and detection loss are put to the channel, 
therefore $\eta$ is the overall transmittance. The value $\eta=10^{-4}$ corresponds to a distance of
120-130km.) The expected value of $n_1+n_m$  is 
$N(1-e^{-\eta\mu_v})\le 10^{-8}N$. Meanwhile, the expected number of dark counts is around $ 10^{-6}N$.
Since the expected number of dark counts there is much larger than the expected number of $n_1+n_m$,
 {\em a little bit} fluctuation of dark counts will totally destroy the estimation
of the value $n_1+n_m$ therefore totally destroy the estimation of $n_1$.
 To make a 
faithful estimation, we request the fluctuation of dark count to be much less than 
the expected value of $n_1+n_m$, $10^{-8}N$, e.g., in the magnitude order of $10^{-9}N$. This is to say,
 one must make sure that the relative fluctuation of dark counts
is less than
$0.1\%$, with a probability {\em exponentially} close to 1 (say, $1-e^{-25}$). This requires $N$ larger than $10^{14}$.
In practice, the system repetition rate is normally less than 80M Hz, i.e., $8\times 10^7/s$. Producing $10^{14}$ 
decoy pulses needs more than
 14 days.    

In the end of their paper\cite{lolo}, they claim that they are able to
show their results with only a few states. This short statement there is rather
vague and by far not a protocol. 
In particular, if they mean the idea in Ref.\cite{lo4}, then it doesn't 
really work as we have shown. If they mean something else and their main claim is based on
that, they should not use the phrase ``for the first time'' in their claim,  
since Ref\cite{lolo} itself is presented latter than our work\cite{wang0}.
We also question their claim to do QKD over 180km. They claimed so
without showing any necessary details. 
{\bf They should at least clearly answer this question: 
 What is the protocol and 
to make a statistically faithful estimation of
lower bound of single photon yield, how many pulses are needed ?} We believe an unreasonably large number
of pulses is needed and one needs more than one month to produce them,
unless they actually use a new method, e.g., our methods reported in Ref\cite{wang0,wang2}
in doing that estimation. Actually, so far they have never considered
the constraint that the number of total pulses in practice can be only reasonably large. In their work, they have
unconsciously assumed exact statistical estimation for small quantities.  

From the methodological viewpoint, in their main protocol, they use the simple-minded method
to solve the joint equations with infinite number of variables, i.e., the yields of each Fock states.
We don't believe that they can really make the long distance QKD with 
only a few different states by the simple-minded mathematical method used in their main protocol.
From their start-point\cite{tot,lolo}, there are infinite variables. It is not surprising to solve the problem
with infinite equations. But the job is non-trivial if there are only a few equations. Therefore,
the authors of Ref\cite{lolo} should clearly state the protocol rather than vaguely claim that they
can make it with a few decoy states in Ref\cite{lolo}.
However, it is possible for anyone to make it with only a few decoy states by a {\em non-trivial} mathematical
method, e.g., the method in our work\cite{wang0}. 
In our work\cite{wang0}, we have put all multi-photon counts into $one$ mixed state, $\rho_c$, therefore,
we $only$ need to consider three variables, the yields of states $|0\rangle\langle 0|,|1\rangle\langle 1|$ and $\rho_c$
with non-trivial $inequalities$. We have chosen reasonable values 
for both $\mu$ and $\mu'$ in using our method\cite{wang0}. If we don't mind decreasing the key 
rate, we can also choose a very small value for one of them. But such a setting is unnecessary since it's key rate
is always lower than the those normal settings. We remind other authors
not to regard a poorly set special case of our method as their own protocol. 

We have shown that, prior to our work\cite{wang0}, none of decoy-state protocol\cite{tot} or idea\cite{lo4} by H. K. Lo et al really works efficiently
in practice.
Actually, so far our result presented in quant-ph/0410075\cite{wang0} is the $unique$ protocol that works efficiently
in practice, by decoy-state method. 
If, in Ref\cite{lolo}, their main claim is actually based on 
something different from their previously announced results\cite{tot,lo4}, since Ref\cite{lolo} itself
is presented later than our work\cite{wang0}, then at least the phrase ``for the first time'' is inappropriate in their claim.

In conclusion, 
the main claim by H.K. Lo et al\cite{lolo} is inappropriate. Their main protocol requires infinite 
number of pulses and their methodology in the main protocol is to straightforwardly solve joint equations
with infinite variables. Their idea in Ref\cite{lo4} doesn't work either because it
implicitly requires at least 14 days to complete one protocol. {\bf If the authors of Ref\cite{lolo} insist
 on their main claim, it should not be difficult for them to answer these simple questions: Prior to our work\cite{wang0}, 
which of their protocol can really work in practice ? If they had one, what is it and where is it ? How many pulses does it need ?} 
We believe that actually, so far our result\cite{wang0} is the {\bf unique} clearly stated protocol that works efficiently
in practice, by decoy-state method. Our method is further developed in Ref\cite{wang2}.
Besides our protocol\cite{wang0}, if anybody happens to know another 
clearly stated decoy-state protocol that works efficiently in practice, please let me know. (email: $wang\_xiangbin@hotmail.com ;
wang@qci.jst.go.jp$) 
\vskip.2in
\noindent
{\bf Note Added:} More than one month after this comment was presented, a separate 
article on decoy-state protocol was presented\cite{loea}. We emphasize that the protocol as stated in 
Ref\cite{loea}, with using the main ideas of our work\cite{wang0}, is {\em different} from Lo's 
earlier idea stated in Ref\cite{lo4}. Their new separate work\cite{loea} does not change the 
fact that prior to our work\cite{wang0}, no decoy-state method can really work efficiently in practice.

The idea stated in\cite{lo4} only suggests watching the counting rates of decoy states of
vacuum
and very weak coherent states and calculating the lower bound
of single photon counts  with these. As we have shown, in this way, the decoy coherent state must be
{\em very weak}: Its average photon number must be less
than the channel transmittance $\eta$ therefore the protocol\cite{lo4} doesn't work due to the statistical
fluctuation of dark count.
However, the method in\cite{loea} suggests watching the counting rates of both
decoy states and signal states and treating them jointly with non-trivial inequalities.
This is indeed the main idea of our method\cite{wang0}. In such a way, the intensity of the decoy coherent state
need not to be {\em very weak}. 

The  difference between their ``Vacuum $+$ Weak decoy coherent state'' protocol\cite{loea} and our protocol\cite{wang0}
is mainly in the specific parameter settings. We have chosen $\mu,\mu'$ in the range of 0.2-0.45 only because we believe
this range gives good results. Our formula for calculation of $\Delta$ also works for the specific parameter setting
used in their work\cite{loea}.  
Definitely we can also use the stronger GLLP formula as recommended in Ref\cite{loea}
for our protocol. Here we suggest using the strongest GLLP formula
given very recently\cite{lost} for key distillation of our protocols\cite{wang0,wang2}. In my opinion,
their result looks more like comparison of different GLLP formulas rather than different decoy-state protocols. 

Our protocol was then improved\cite{wang2}.  We believe that the key rate of the protocol in Ref\cite{wang2} is quite good
even compared with their new work\cite{loea}, using the same GLLP formula for key distillation.  


\begin{references}
\bibitem{lolo} H. K. Lo et al, quant-ph/0411004.
\bibitem{tot} H.-K. Lo et al, http://www.fields.utoronto.ca/programs /scientific/04-05/quantumIC/abstracts/lo.ppt;
/lo.pdf : Decoy state quantum key distribution (QKD), page 27. And also: page 18, 
http://www.newton.cam.ac.uk/webseminars/pg+ws/2004 /qisw01/0826/lo/ 
\bibitem{lo4} H.-K. Lo, p.17, Proceedings of 2004 IEEE Int. Symp. on Inf. Theor., Hune 27-July 2, 2004, Chicago.
\bibitem{wang2} X. B. Wang, quant-ph/0411047, v5, Feb 21, 2005; and v1-v3, 
2004. (Note: v4 is a wrong file which is identical to quant-ph/0410075).
\bibitem{wang0} X. B. Wang, quant-ph/0410075.
\bibitem{loea}X.Ma, B. Qi, Yi Zhao and H.-K. Lo,
quant-ph/0503005,  March 1, 2005.  
\bibitem{lost}H. K. Lo, quant-ph/0503004.
\end{references}
\end{document}